\documentclass[%
 reprint,
superscriptaddress,
 amsmath,amssymb,
 aps,
prb,
]{revtex4-2}

\usepackage{graphicx}
\usepackage{dcolumn}
\usepackage{bm}
\usepackage{xcolor}

\begin{document}

\preprint{APS/123-QED}

\title{Ab initio molecular dynamics modelling of \\
organic crystals electro-optical properties}

\author{Hovan Lee}
\affiliation{%
Department of Physics, Faculty of Natural \& Mathematical Sciences, King's College London, London, WC2R2LS, UK
}%

\author{Rui Hou}
\affiliation{Department of Physics, Hong Kong University of Science and Technology, Hong Kong, China}

\author{Ding Pan}
\affiliation{Department of Physics and Department of Chemistry, Hong Kong University of Science and Technology, Hong Kong, China}
\affiliation{HKUST Fok Ying Tung Research Institute, Guangzhou, China}

\author{Mostafa Shalaby}
\affiliation{3 Swiss Terahertz Research-Zurich, Technopark, 8005 Zurich, Switzerland and Park Innovaare, 5234
Villigen, Switzerland}
\affiliation{Key Laboratory of Terahertz Optoelectronics, Beijing Advanced Innovation Center for Imaging Technology CNU, Beijing 100048, China}

\author{Cedric Weber}%
\affiliation{%
Department of Physics, Faculty of Natural \& Mathematical Sciences, King's College London, London, WC2R2LS, UK
}%




\date{\today}

\begin{abstract}
Molecular dynamics calculations were preformed on organic crystals 4-N,N-dimethylamino-4'-N'-methyl-stilbazolium tosylate (DAST) and 4-N,N-dimethylamino-4'-N'-methylstilbazolium 2,4,6-trimethylbenzenesulfonate (DSTMS). Vibrational modes of the structures were investigated to examine the single unit cell phononic contribution of the organic crystals to their terahertz generating capabilities. Linear optical properties were also calculated from snapshots of the molecular dynamics structures through Kubo-Greenwood relations, and compared with experimental transmission.
\end{abstract}

\maketitle


\section{\label{sec:level1}Introduction}


Throughout the past decades, the research on terahertz (0.1 - 10 THz) frequency generation have been motivated by the plethora of applications of THz technology \cite{THzSpectroscopyMasayoshi:2007, THzSpectroscopyJepsen:2011,THzSpectroscopyBaxter:2011}; spanning from biomedical purposes \cite{ApplicationYang:2016,biomed1,biomed2,biomed3} to next generation wireless communications \cite{next_gen1,next_gen2,next_gen3}, amongst others \cite{ApplicationUlbricht:2011,ApplicationRazzari:2009,ApplicationGreenland:2010,ApplicationShalaby:2017,ApplicationKolner:2008,ApplicationMcIntosh}. The development of these applications were built upon numerous technological advances such as the advent of quantum cascade lasers, femtosecond lasers, time domain spectroscopy and various THz generating sources. Of these THz sources, there exists a general preference across most areas of discipline for broad band, highly efficient and convenient methods of THz generation, often due to the versatility of non-ionising and non-invasive nature of the radiation.

On this note, table-top organic crystal THz sources have become popularised due to their portability and adaptability, as these crystals are capable of generating THz radiation with efficiencies of
$\sim 1\%$
when pumped with the conventional near infrared Ti:Sapphire laser. Moreover, these crystals offer varying applicable pump wavelengths and THz emission spectra, granting the user a method of altering the THz bandwidth, as the need arises.

These organic crystals, such as 4-N,N-dimethylamino-4'-N'-methyl-stilbazolium tosylate (DAST) and 4-N,N-dimethylamino-4'-N'-methylstilbazolium 2,4,6-trimethylbenzenesulfonate (DSTMS), generate THz radiation through the nonlinear optical process of electro-optic rectification. In this process the electronic structure of the material favours a preferred direction of polarisation when influenced by an electromagnetic field. In DAST and DSTMS, this attribute comes from the charge transfer between donors and acceptors, conducted via $\pi$-electrons \cite{pi-electrons}, which are abundant in the conjugated chains of their respective structures. Through this mechanism, the time-confined (and hence wide spectral bandwidth) nature of the femtosecond pump pulse induces a frequency dependent polarisation mixing, the beating of this polarisation results in THz emission.

Vibrational excitations can also influence the polarisation along these conjugated chains \cite{conjugated-polarisation}; vibrational modes distort the ionic structure of the chains, and in turn, the electronic charge conforms to the distorted structure. This can cause further changes of charge distribution along the chain, altering the overall polarisation and polarisability of the materials.

\begin{figure}
    \centering
    \includegraphics[width=0.6\linewidth]{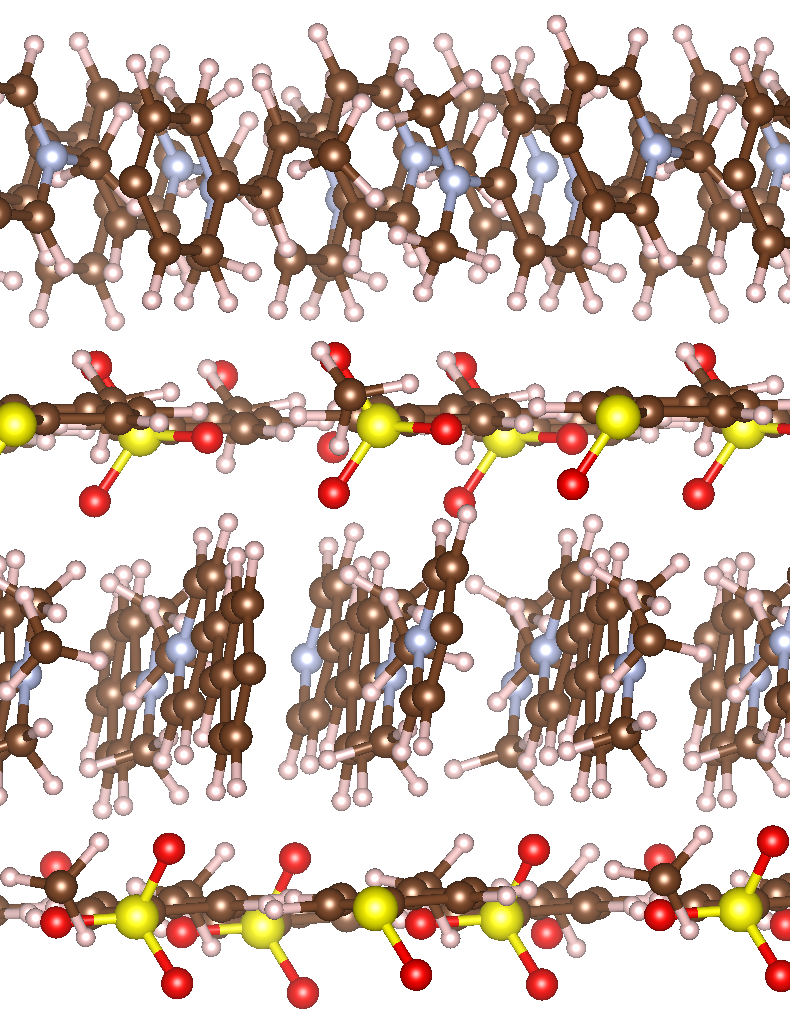}
    \caption{Crystal packing structure of organic crystal DAST.}
    \label{dast_packing}
\end{figure}

Phononic contributions on the linear electro-optic response of DAST has already been measured \cite{phonon_exp} and also calculated for a single anion-cation pair, in terms of effects on the near infrared Fourier transform Raman, and Fourier transform infrared spectra \cite{phonon_coupling}. However, such calculations have not been performed on the full crystal packing structure of either DAST nor DSTMS. That is to say that the previous works assume periodicity of the polarisation of a single anion-cation pair to be replicated across the full crystal. This does not account for more complex vibrational modes, such as neighbouring conjugated chains oscillating out of phase, or misalignment between successive anion-cation layers (as shown in Fig.\ref{dast_packing} for DAST). These effects expand upon the periodicity of a single anion-cation pair, allowing higher orders of variations of polarisability, which influences the calculated electro-optic response of the material.

Furthermore, it was shown \cite{bosshard} that vibrational modes account for roughly $25\%$ of the linear electro-optic coefficient in DAST. Therefore it is essential, even on the linear response level, to include vibrational excitations in the calculation of the optical properties of these organic crystals.

Moreover, little has been done on understanding what roles these vibrational excitations play in the absorption and transmission of THz radiation. Considering that the THz frequency range correspond to vibrational and rotational excitation modes in materials, it is quintessential to develop methods of modelling and understanding absorption and transmission separately. Traditionally, phonon modes in first-principle methods are calculated via the dynamical matrix, which requires double partial differentials of the system energy with respect to spatial coordinates for each atom in the system. This is computationally costly for the $>200$ atom crystal packing unit cells of DAST and DSTMS.

Therefore in this work, we review a method to model optical transmission and absorption in relation to the lattice dynamics; we perform molecular dynamics simulations on the DAST and DSTMS crystal packing structures, analyse the Fourier transform of ionic displacements, and utilise Kubo-Greenwood relations to compare our results to experimentally obtained DAST transmission spectra.

\section{\label{sec:level1}Theory}

\begin{figure*}
    \centering
    \includegraphics[width=0.7\linewidth]{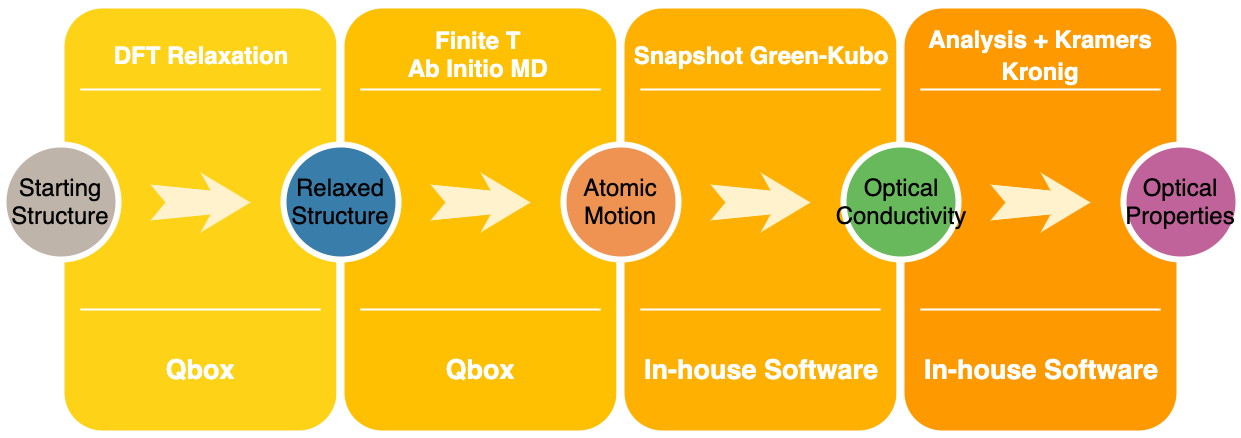}
    \caption{Flowchart of the procedure of calculations in obtaining the optical properties of the simulated materials.}
    \label{scheme}
\end{figure*}

Molecular dynamics (MD) calculations were performed using the first-principles planewave software Qbox. These calculations were carried out,
at ambient temperature
using the B3LYP hybrid density functional, for the DAST and DSTMS crystal packing structures.
Atomic coordinates at time step intervals of the MD simulation were obtained, and their optical conductivities were calculated through Kubo-Greenwood relations using in-house software:
\begin{widetext}
\begin{equation}
    \Re(\sigma^{\alpha\beta}(\nu))=\frac{2\pi e^2 \hbar}{V}\sum_k\int d\omega'\frac{f(\omega')-f(\omega'+\nu)}{\nu} tr\{\upsilon_\alpha(\mathbf{k})A(\mathbf{k},\omega')\upsilon_\beta(\mathbf{k})A(\mathbf{k},\omega'+\nu)\}
\end{equation}
\end{widetext}
where $\Re(\sigma^{\alpha\beta}(\nu))$ signifies the real part of the conductivity matrix element in the Cartesian directions $\alpha$ and $\beta$ at frequency $\upsilon$, $e$ is the magnitude of the electron charge, $V$ is the volume of the simulation, $f(\omega')$ is the Fermi-Dirac distribution at energy $\omega'$, and $tr\{\dots\}$ denotes the trace over spectral functions in Kohn-Sham basis $A(\mathbf{k},\omega)$ and Fermi velocities $\upsilon_\alpha(\mathbf{k})$. The upper limit of the integral over energy $\omega'$ was taken such that all relevant optical transitions are included in the calculation.

The Kubo-Greenwood relations were preferred over the classical Green-Kubo relations (as applied in works such as \cite{ding1,ding2}) due to the large unit cells of the crystals; a integration of dipole autocorrelation functions over time, as is necessary in classical Green-Kubo, would be inhibitively expensive. In this work, we assume that the electronic states relax instantaneously compared to ionic motion, allowing us to utilise the quantum analogue of integration of Kohn-Sham basis Fermi-velocity operators over energy as an estimation of linear optical response. Other examples of the same technique are available in \cite{cedric_hemoglobin,cedric_myo}.

The optical conductivity ($\pmb{\sigma}$) of a material can be expressed as a frequency dependant rank 2 tensor, and is related to the complex dielectric tensor of a material:
\begin{equation}
    \pmb{\epsilon}_{complex}(\omega)=\pmb{\epsilon}^{(1)}(\omega)+\frac{4\pi i}{\omega}\pmb{\sigma}(\omega)=\pmb{\epsilon}^{(1)}+i\pmb{\epsilon}^{(2)}
\end{equation}
where $\pmb{\epsilon}^{(1)}$ and $\pmb{\epsilon}^{(2)}$ are the real and imaginary parts of the complex dielectric functional respectively. $\pmb{\epsilon}^{(1)}$ tensor elements were obtained through Kramers Kronig relations\cite{kk}:
\begin{equation}
    \epsilon_{\alpha\beta}^{(1)}(\omega)=1+\frac{2}{\pi}\text{P}\int_0^\infty\frac{\epsilon_{\alpha\beta}^{(2)}(\omega')\omega'}{\omega'^2-\omega^2}d\omega'
\end{equation}
where P denotes the principle value.

The dielectric tensor of the geometry optimised structure was diagonalised to obtain the equilibrium dielectric functions along the crystal axes of the material, together with the rotational vectors which transforms the system into the crystal frame. The dielectric tensors of the MD snapshots were then transformed with respect to these equilibrium frame rotational vectors, in order to calculate the optical properties along the axes of the crystal. The dielectric function along the $a$-axis of the crystal, which contains the largest nonlinear coefficient $d_{111}$\cite{d111} of DAST and DSTMS, was geometrically identified.

The optical constants of the solid\cite{Dresselhaus_solidstate} are expressed as:
\begin{equation}
    \sqrt{\epsilon_{complex}^{11}(\omega)}=\tilde{n}(\omega)+i\tilde{k}(\omega)
\end{equation}
where $\tilde{n}$ is the refractive index and $\tilde{k}$ is the extinction coefficient. From these terms, the absorption coefficient was calculated:
\begin{equation}
    \alpha(\omega)=\frac{2\omega\tilde{k}(\omega)}{c}
\end{equation}
here $\alpha$ is the absorption coefficient, and $c$ is the speed of light. Which was used to calculate the absorption of the material of length $l$:
\begin{equation}
    \mathcal{A}=exp(-\alpha l)
\end{equation}
and the transmission of the material is then:
\begin{equation}
    \mathcal{T}=1-\mathcal{A}-\mathcal{R}=1-exp(-\alpha l)-\frac{(1-\tilde{n})^2+\tilde{k}^2}{(1+\tilde{n})^2+\tilde{k}^2}
\end{equation}
where $\mathcal{R}$ is the reflectivity of the material.

The full procedure of calculating the optical properties in DAST and DSTMS, as carried out in this work, is illustrated in Fig.\ref{scheme}.

\section{\label{sec:level1}Results}
\begin{figure}
    \centering
    \includegraphics[width=\linewidth]{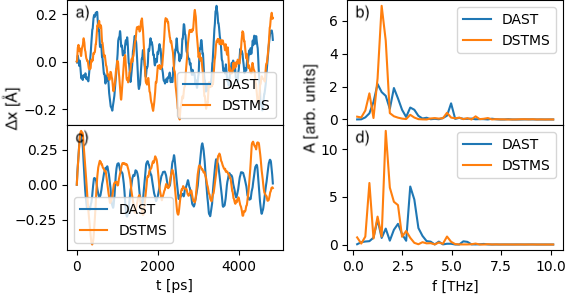}
    \caption{a) Difference from equilibrium value of centre of mass between the centre of mass of successive anion chains along $a$-axis for DAST and DSTMS. b) Normalised amplitude of Fourier transform of distance from equilibrium along $a$-axis. c) Difference from equilibrium value of centre of mass between the centre of mass of successive anion chains along $b$-axis. d) Normalised amplitude of Fourier transform of distance from equilibrium along $b$-axis. Both sets of Fourier transforms are normalised to the same value, this permits comparisons of oscillation amplitudes across different crystals and different crystal axes.}
    \label{diff}
\end{figure}

The centre of mass of  the stibazolium chains were calculated, and their change in displacement along the crystal a-axis between 2 neighbouring stibazolium chains is shown in Fig.\ref{diff}.a) for DAST and DSTMS at $300$ $K$. In this figure, we observe oscillations around the equilibrium distance, with amplitudes of $\sim 0.2 \AA$ for both organic crystals, showcasing effects that cannot be observed from simulating a single anion-cation pair alone. Moreover, the amplitude of Fourier transforms of the oscillations are shown in Fig.\ref{diff}.b), here we observe the largest oscillations in the ranges of $0-4$ $THz$ for DAST, and $0-2$ $THz$ for DSTMS. The peak spectral amplitude of DSTMS is roughly threefold of the peak amplitude of DAST, suggesting that at room temperature for the oscillation mode at $3$ $THz$, the contribution of vibrational excitations on the linear response level is even greater than that of DAST. 

In Fig.\ref{diff}.c) and Fig.\ref{diff}.d) these effects are shown, for the crystal $b$-axis, on a greater extent: the waveform of the oscillations are smoother and at greater amplitudes, at $\sim 0.4 \AA$. In the Fourier analysis, where the amplitudes are normalised to the same value as in Fig.\ref{diff}.a), the band of oscillation frequencies for DAST remains between $0-4$ $THz$, with a twofold increased relative amplitude in the range of $3-4$ $THz$, where as the DSTMS band increases in width from $1-2$ $THz$ to $0-3$ $THz$, whilst maintaining similar frequency features at double the amplitude between the $a$-axis and $b-axis$. This implies that the effects of vibrational excitations are greater along the $b$-axis when compared to the $a$-axis.

This highlights the importance of accounting for structural vibrations when examining the THz optical properties of DAST and DSTMS; the charge transfer axes of the stibazolium conjugated chains are at angles $\pm23^{\circ}$ with respect to the $a$-axis on the $a$-$b$ plane. These oscillations correspond to geometric changes of the stibazolium chains, leading to changes in the charge distribution and polarisability of the crystal structure on the $a$-$b$ plane. Thus altering the non linear optical effects of both DAST and DSTMS.

\begin{figure}
    \centering
    \includegraphics[width=1\linewidth]{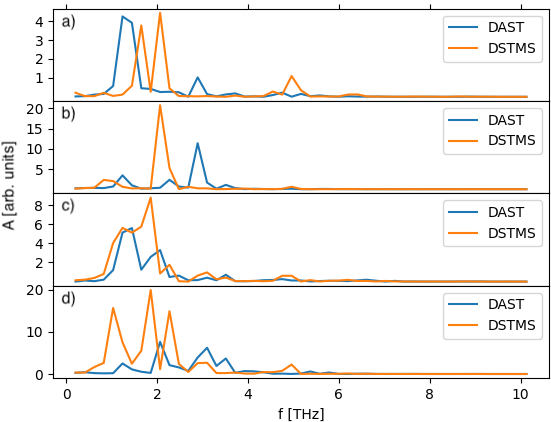}
    \caption{Fourier transform of the relative displacements of the centre of mass of neighbouring cations in DAST and DSTMS, calculated through molecular dynamics with an ionic temperatures of $150$ $K$, shown in a) for displacements along the $a$-axis and b) along the $b$-axis, and $450$ $K$, shown in c) for displacements along the $a$-axis and d) along the $b$-axis.}
    \label{temp}
\end{figure}

The MD calculations were also performed for both crystals at $150$ $K$ and $450$ $K$, and the Fourier transform of stibazolium chain displacements, with the same normalisation value compared to Fig.\ref{diff}, are shown in Fig.\ref{temp}. Fig.\ref{temp}.a) showcases the Fourier transforms of the displacements, parallel to the $a$-axis, between successive stibazolium cations at $150$ $K$. With comparisons to the $300$ $K$ DAST simulation, the spectral bandwidth of the oscillations decreased to the range of $1-2$ $THz$, concomitant with a doubling of the peak modal amplitude. Meanwhile for DSTMS, we observe a split in the $300$ $K$, $1-2$ $THz$ band into two separate modes at $1.5$ $THz$ and $2$ $THz$. 

The $b$-axis equivalent of Fig.\ref{temp}.a) is shown in Fig.\ref{temp}.b). Compared to Fig.\ref{diff}.d), the $150$ $K$ simulations for both DAST and DSTMS exhibit similar peaks at $1.3$ $THz$, $2.2$ $THz$ and $3$ $THz$ for DAST and at $1$ $THz$ and $2$ $THz$ for DSTMS. These peaks are both more narrow and have higher amplitudes compared to their $300$ $K$ counterparts.

In our view, the changes in both Fourier spectra in comparison to their $300$ $K$ counterparts are due to the freeze out of vibrational modes as temperature decreases, therefore increasing the amplitude of the fundamental vibrational modes that are not frozen. This suggests that the contribution of vibrational excitations are relevant when calculating the electro-optic response of these organic crystals at a temperature of $150$ $K$.

The $450$ $K$ MD simulation results are shown in Fig.\ref{temp}.c) for the $a$-axis and Fig.\ref{temp}.d) for the $b$-axis. For both crystals and in both axes, we observe a general trend of a broadening in spectral features and an increase in modal amplitude when compared to the $300$ $K$ simulation, without the onset of additional vibrational modes.

These results suggest that an increase of temperature from $300$ $K$ to $450$ $K$ does not unfreeze any additional vibrational modes. The effect of temperature is only observed as an increase in vibrational amplitude and the increase in vibrational bandwidth.

\begin{figure}
    \centering
    \includegraphics[width=0.9\linewidth]{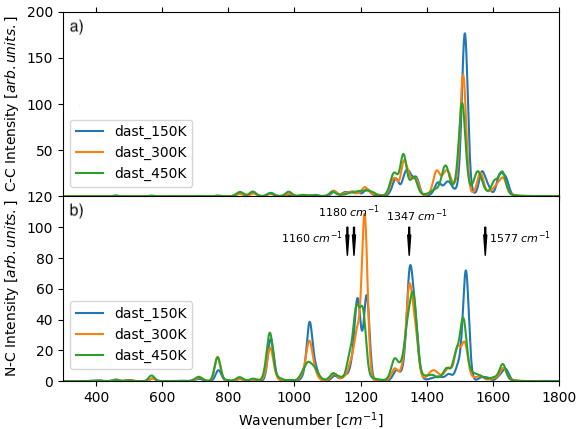}
    \caption{Top: The Gaussian broadened, simulated bond length vibrational spectra calculated through the autocorrelation function of the carbon carbon bond lengths of the conjugated cation chain of the molecular dynamics simulation at different temperatures. Bottom: Gaussian broadened, simulated bond length vibrational spectra of the nitrogen carbon bond lengths of the cation chain. Additional black arrows indicate strongest modes of the infrared and Raman spectra as experimentally observed in \cite{bosshard}.}
    \label{autocorr}
\end{figure}


In Fig.\ref{autocorr}.a) the MD bond lengths between carbon atoms of two (consecutive and disaligned) layers of conjugated cationic chains were analysed and displayed as vibrational spectra, through the use of the autocorrelation function and Gaussian broadening, at $150$ $K$, $300$ $K$ and $450$ $K$. The same technique was also utilised in studying the chemical speciation of hydrocarbons\cite{hou2021raman}. At all temperatures, an identifiable peak at $1500$ $cm^{-1}$ is observed, this peak decreases in amplitude and increases in spectral bandwidth as temperature increases, mirroring our observations between Fig.\ref{diff} and Fig.\ref{temp}. 

The nitrogen-carbon bond lengths at the ends of the conjugated cationic chains were also examined in Fig.\ref{autocorr}.b). Here we do not observe the same trend of peak broadening as temperature increases, instead the order of increasing amplitude seems to change at different wavenumbers. 
We attribute this to the assimilation of the two features observed $\sim 1200$ $nm$ at a temperature of $150$ $K$, leading to first an increase in amplitude at $300$ $K$, and subsequently a spectral broadening at $450$ $K$.
Nevertheless, we observe similarities between these bond length vibrations, and the infrared (IR) and Raman active modes as experimentally measured in \cite{bosshard}, which are depicted as black arrows in Fig.\ref{autocorr}.b). For the IR and Raman mode at $1347$ $cm^{-1}$, we observe a match in our spectra, and therefore assign this mode to the nitrogen-carbon bond length vibrations. We also observe a mode at $\sim 1200$ $cm^{-1}$, which is close in wavenumber to the IR and Raman modes at $1160$ amd $1180$ $cm^{-1}$. Lastly, we attribute the IR and Raman active mode of $1577$ $cm^{-1}$ to vibrational modes that are not considered in this analysis, namely the change in bond angles and the vibrations of the anions.

\begin{figure}
    \centering
    \includegraphics[width=\linewidth]{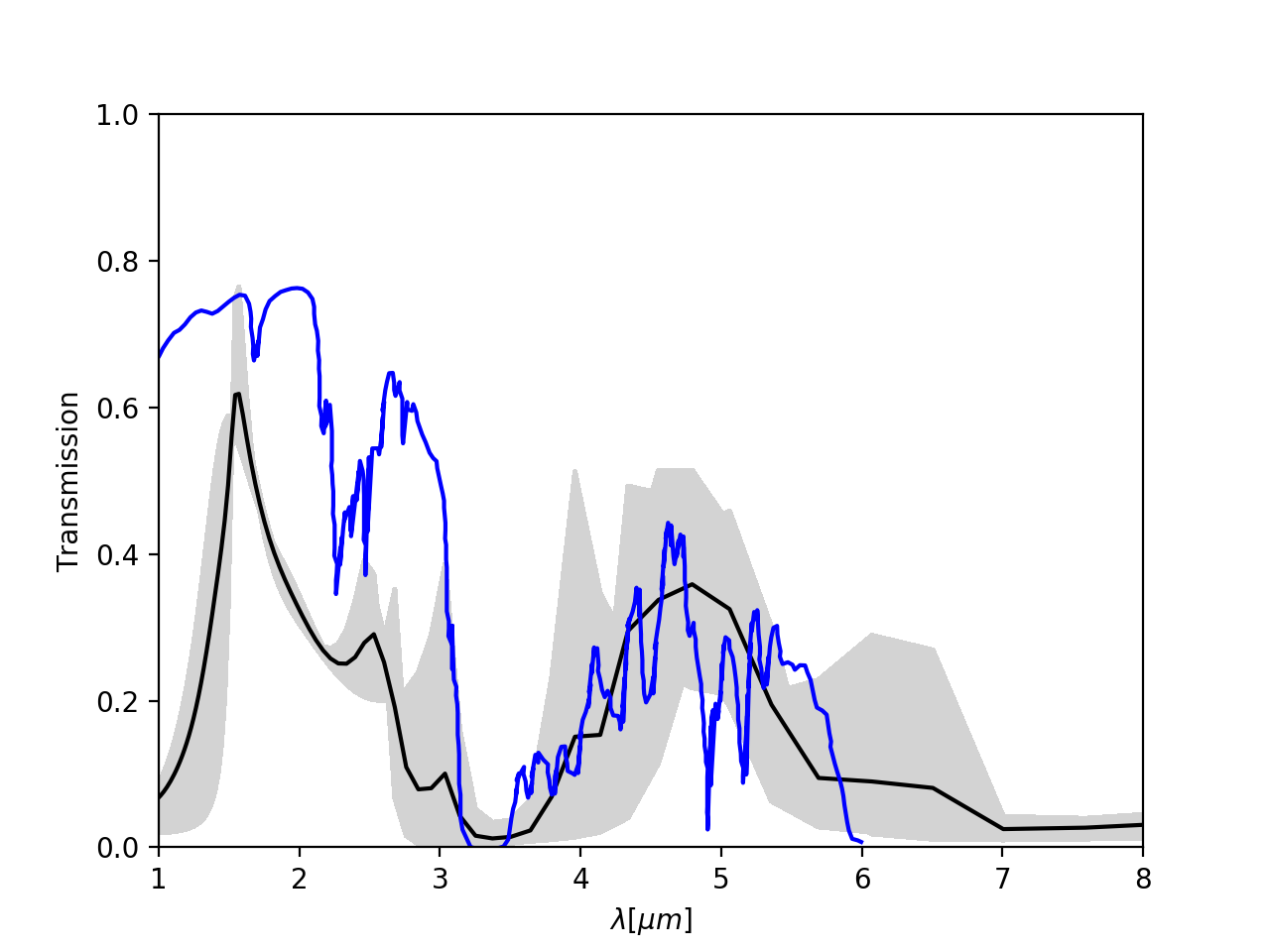}
    \caption{Comparison of calculated time-step averaged transmission of DAST in black, variance in grey, and the experimentally obtained spectra \cite{exp} in blue.}
    \label{transmission}
\end{figure}

The transmission of DAST was calculated through Kubo-Greenwood relations from a series of MD snapshots and is shown in Fig.\ref{transmission}, where the MD time step average transmission is shown in black and the variance of the transmission is shown in grey. The experimentally obtained transmission spectra \cite{exp} of DAST is also shown in blue as a comparison.

We observe matching features in both the experimental and calculated transmission at $\sim1$ $THz$, $2.5$ $THz$ and $4.5$ $THz$, and troughs at $\sim2.2$ $THz$ and $3.5$ $THz$. The similarity in line shape suggest that, as opposed to THz generation, nonlinear corrections may not play a significant role in the optical transmission of DAST.

\section{\label{sec:level1}Conclusion}

In this work, we investigated a computationally inexpensive method to calculate the THz optical properties of DAST and DSTMS through molecular dynamics and linear Kubo-Greenwood relations. This same technique could be extended to designing new materials, in which the electro-optic response of the material could be calculated at a relatively low computational cost.

Through the analysis of the relative displacements of neighbouring cation centres of mass at various temperatures, we were able to identify a freeze out in vibrational modes when decreasing the molecular dynamics simulation temperature from $300$ $K$ to $150$ $K$, whilst no additional modes were observed at $450$ $K$. We also observed that the modal amplitudes of DSTMS cation oscillations are generally larger than those of DAST, this suggests that the DSTMS linear electro-optic response has a higher vibrational component when compared to DAST.

By considering the molecular dynamics vibrations of carbon-carbon and nitrogen-carbon bonds, we observe vibrational modes comparable to experimentally observed infrared (IR) and Raman spectra. We attribute the IR and Raman mode of $1347$ $cm^{-1}$ to nitrogen-carbon bond length oscillations.

The transmission of DAST was also calculated for various time steps in the molecular dynamics simulation through Kubo-Greenwood relations. The calculated spectrum was found to contain similar features compared to experimentally obtained results.

\begin{acknowledgments}
CW was supported by grant EP/R02992X/1 from the UK Engineering and Physical Sciences Research Council (EPSRC). D.P. acknowledges support from the Croucher Foundation through the Croucher Innovation Award, and
National Natural Science Foundation of China through the Excellent Young Scientists Fund.
\end{acknowledgments}

\appendix

\bibliography{apssamp}

\end{document}